\documentclass{ws-procs9x6}
\usepackage{epsfig}
\usepackage{wrapfig}
\usepackage{subfigure}

\newcommand{\qsq}      {\ensuremath{Q^2}}
\newcommand{\ttra}     {\ensuremath{|t|}}

\newcommand{\gev}      {\ensuremath{\rm GeV}}
\newcommand{\gevsq}    {\ensuremath{\rm GeV^2}}

\newcommand{\rhoz}     {\ensuremath{\rho^{0}}}

\newcommand{\jpsi}     {\ensuremath{J/\psi}}

\newcommand{\qqbar}    {\ensuremath{q\bar{q}}}

\newcommand{\Pom}      {\ensuremath{I\!\!P}}              
\newcommand{\pom}      {\ensuremath{I\!\!P}}

\newcommand{\apom}     {\ensuremath{\alpha_{\Pom}}}

\newcommand{\aprim} {\ensuremath{\alpha {'}}}

\newcommand{\rzqzz}    {\ensuremath{r_{00}^{04}}}

\begin{document}
\title{Summary of the ``Diffraction and Vector Mesons'' working group
at DIS06}

\author{H. Lim}
  \address{Agonne Natonal Laboratory, Argonne, Illinois 60439-4815, USA}

\author{L. Schoeffel}
  \address{DAPNIA/Service de physique des particules, CEA/Saclay, 91191
Gif-sur-Yvette cedex, France}

\author{M. Strikman}
\address{ Penn State University, University Park,
PA 16801, PA, USA}

\maketitle

\abstracts{
We survey the contributions presented in the working group ``Diffraction
and Vector Mesons''
at the XIV International Workshop on Deep Inelastic Scattering.
}

\maketitle
\section{Introduction}
Studies of diffractive processes provide a much more detailed information about the pattern of the high energy strong interactions than inclusive cross section. One can infer from these studies both the transverse range of the interaction  as well as its intensity.
The knowledge of the strength of the interaction as a function of the impact parameter provides a direct information on  a possible  proximity of interaction of hadrons or small dipoles 
with nucleons to the  black disk limit (BDL)  of  the maximal strength of the interaction.   The BDL is especially interesting for the case of the interaction of the small objects as it corresponds to the situation when the strong coupling constant is small but the expansion over the twists breaks down.

%-------------------------------------------------

\section{Diffraction and factorization}

%-------------------------------------------------
In hadron-hadron (or lepton-hadron) scattering a substantial fraction of the total cross
section is due to diffractive reactions.
In elastic $p\bar{p}$ scattering,
both projectiles emerge intact in the final state,  whereas
single or double diffractive dissociation corresponds to one or both
of them being scattered into  a resonance or
continuum state with mass, $M_X\ll \sqrt{s}$.  The two groups
of final-state particles are well separated in phase space and in
particular have a large gap in rapidity between them. 
Similar features
hold for lepton-hadron scattering with $l+p \rightarrow l+X+Y$ observed at HERA,
the  low-mass excited state $Y$ carrying the same quantum numbers
 (except spin)
  as the incident proton.

Then, soft diffractive interactions follow specific properties which 
find a natural 
description 
within Regge
theory, in which they can be interpreted as 
mediated by the leading singularity in the angular momentum plane with vacuum quantum numbers, called Pomeron. 

In the perturbative regime Pomeron like behavior of the elastic amplitude emerges at high energies due to the  exchange of a ladder of gluons and quarks in t-channel.  In difference from the {\it universal}  Pomeron of the Regge theory  perturbative Pomeron is a resolution dependent construction which is not directly connected to the nonperturbative Pomeron.

\begin{wrapfigure}{r}{5cm}
 \includegraphics[totalheight=4cm]{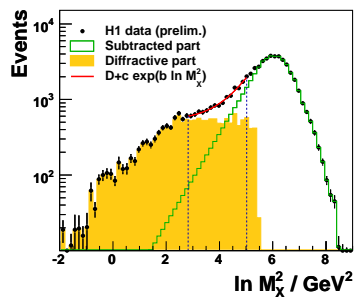}
\caption{ $\ln M^2_X $ spectra with the different contributions : diffractive plateau (full histogram) and high mass peak due to non-diffractive events.}
\label{Fig3i}
\end{wrapfigure}
Following the above considerations, it is clear that diffractive events (at HERA) 
can be selected by (at least) three methods: (a) by the requirement of the existence of a large
rapidity gap (LRG) between the produced hadronic system X and the outgoing
hadron, (b) by tagging leading baryons which carry a large
fraction of the incoming hadron beam energy, and (c) using 
the $M_{X}$  method which is
based on the different characteristics of the $M_{X}$ distributions in 
diffractive and non-diffractive processes according to the Regge model (see Fig.~\ref{Fig3i}). 
 Using this model for hard processes which are superposition of interactions with hard and soft Pomeron may potentially introduce an uncertainty in the analysis.
 Also, LRG method and proton tagging are commonly used at the TEVATRON to characterize
a diffractive event.

Note that samples defined in these ways also contain processes 
for which  nondiffractive mechanisms become important which one can try to model via contribution of subleading Regge trajectories, while the $M_{X}$ method suppresses their contributions. Also, tagging the outgoing hadron presents the interesting
advantage of excluding background from
dissociative processes, $l+p \to l+X+Y$, with the drawback that the acceptance
and then kinematic coverage are quite limited for this technique.

Following arguments related to the color transparency phenomenon in QCD, one can prove the QCD factorization theorem for semi inclusive processes $l+p\to l+X+h$ for fixed Feynman $x_L\equiv  1-x_{\pom}$ and transverse momentum  of the hadron \cite{Collins:1997sr}. In most of the data analyses this theorem is supplemented by an assumption motivated by the Regge model that the
diffractive (for small x) parton densities (DPDFs) can be factorized as a product of the function of $x_{\pom}, p_t$ and a function of $\beta=x/x_{\pom}$, $Q^2$.
Then, diffractive parton density functions (DPDFs)  integrated over $p_t$
can be extracted from the measurements of diffractive structure function ($F_2^D$) and
can be used to predict the rates of diffractive jet 
or diffractive heavy flavour 
processes. Though the DPDFs extracted from HERA data overestimate the rate
of diffractive dijet process at the TEVATRON, it could be explained by 
absorptive effects since the interaction of protons at small impact parameters is practically black and cannot  lead to diffraction. Similar suppression effect is an essential ingredient of the preditions on exclusive diffractive Higgs production at the LHC.

%-------------------------------------------------

\subsection{Leading neutron measurement} 

%-------------------------------------------------

\begin{figure}
  \hspace{2mm}
  \includegraphics[height=.27\textheight]{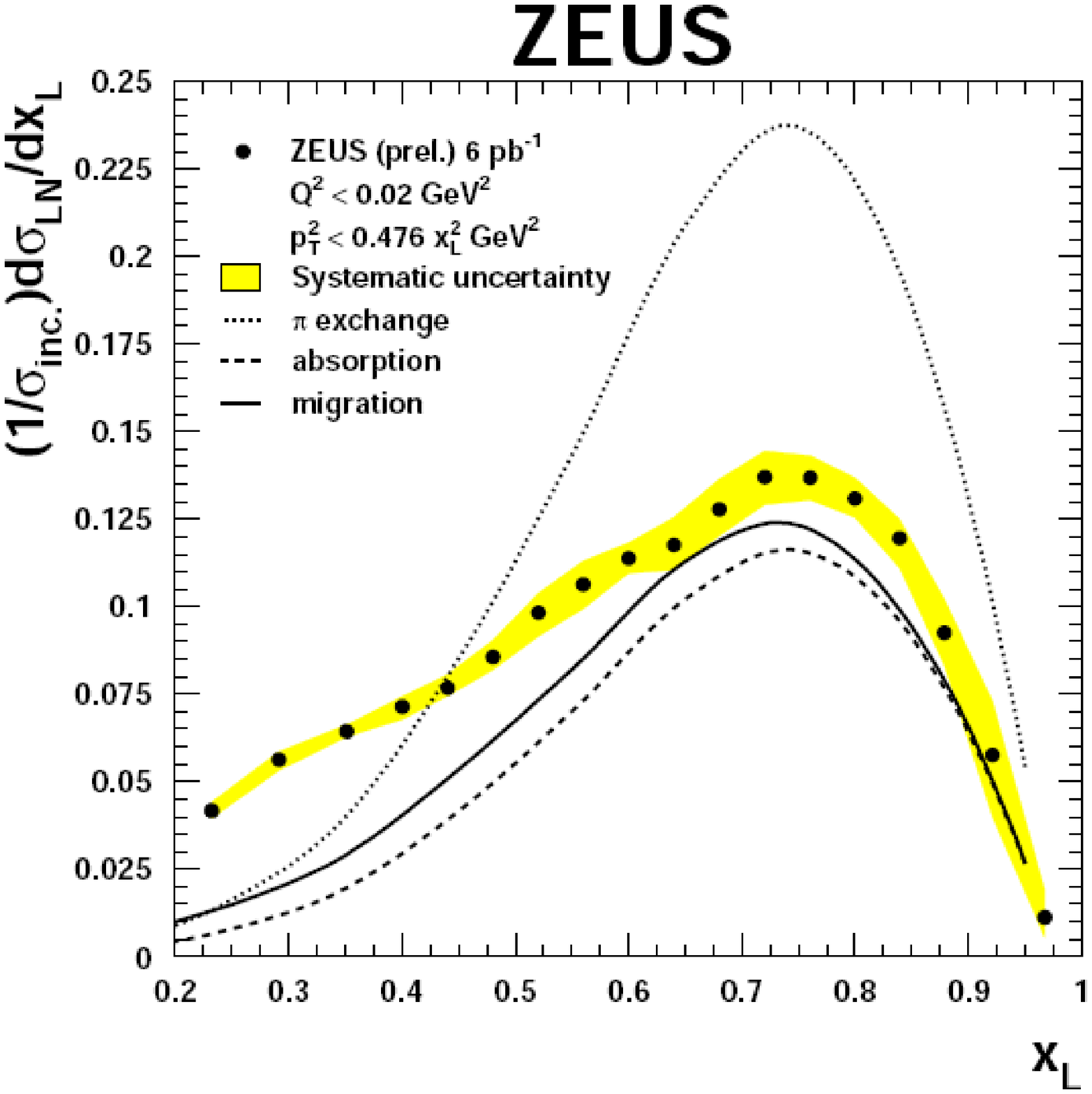}
  \hspace{7mm}
  \includegraphics[height=.27\textheight]{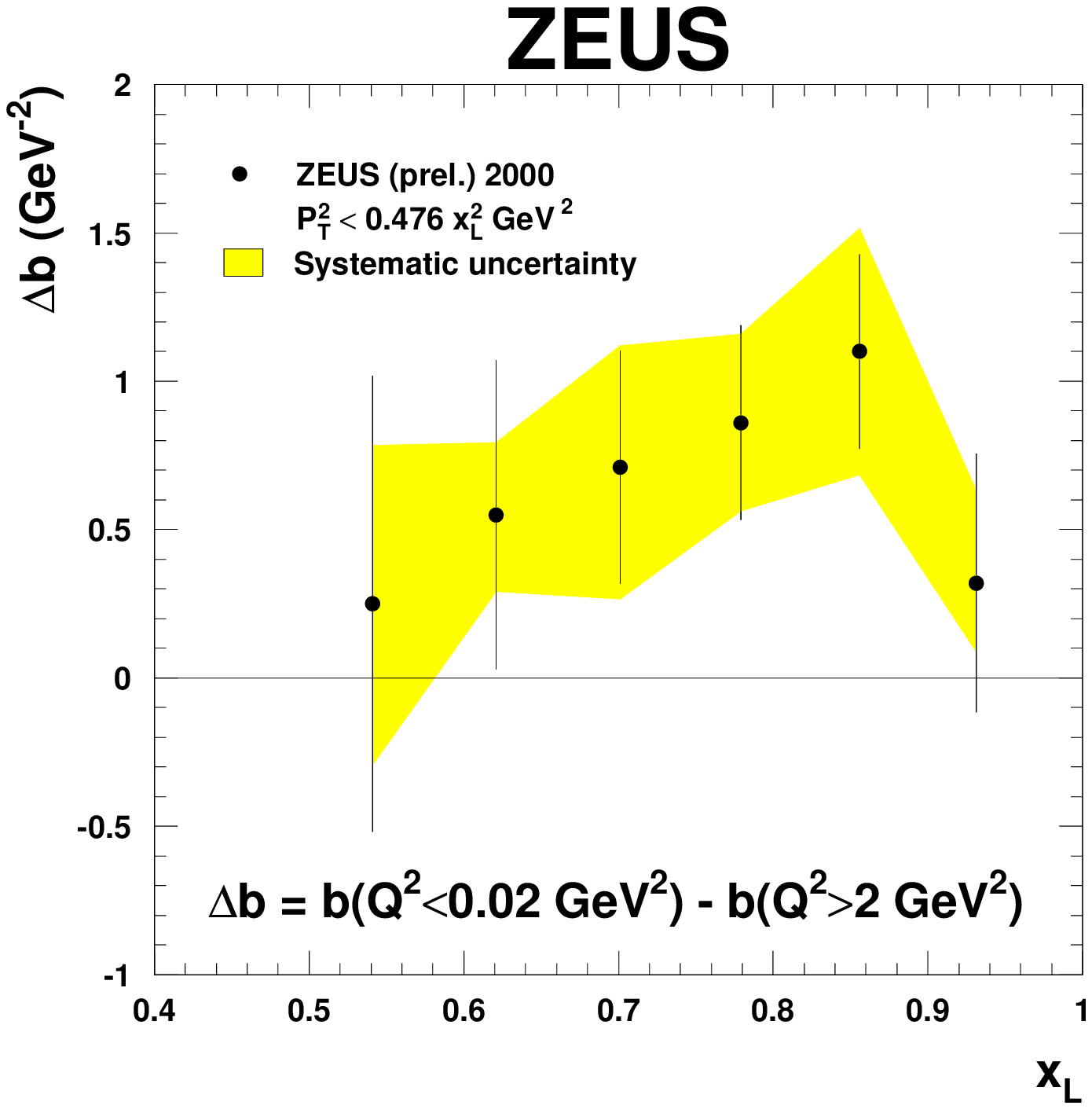}
  \caption{(left) Leading neutron energy spectra for the photoproduction 
	sample. Curves are from the predictions from the one-pion exchange
	model including the effects of neutron absorption.
	(right) The difference between
	slopes in photoproduction and DIS vesus $x_{L}$.}
  \label{fig:LN_fig}
\end{figure}

The production of leading baryons in $ep$ scattering has been studied
at the HERA collider. The ZEUS Collaboration has presented recent
measurement of the energy distributions of leading neutrons in $ep$
by real and virtual photons~\cite{mara} for which in the DIS limit the QCD factorization theorem\cite{Collins:1997sr} should be valid. The ZEUS forward neutron
calorimeter (FNC) measures neutrons which carry a large fraction, $x_L$ of the
incoming proton beam energy and are produced at a small angle with
respect to the incoming proton direction. Measurements are consistent with the expectation that multiPomeron exchanges which lead to absorption effects
decrease with increase of $Q^2$. Data also indicate that the neutron multiplicities are  pretty similar for the case of the processes dominated by scattering off quarks and off gluons (dijet production) which is consistent with the fragmentation scenario of 
Ref.~\cite{Frankfurt:1997ij}. 

The DIS data were described in the one-pion exchange model including effects of absorption 
~\cite{KKMR}  (see Fig.~\ref{fig:LN_fig} (left)). 
The $p_{T} ^{2}$ distributions of neutrons were parametrized using
$\exp(-b p_{T} ^2)$. 
Though the tendency of $b$ as a function of $x_L$ is similar between
the theoretical predictions and data,  none of the models describes the data
in scale.   In Fig. \ref{fig:LN_fig} (right), the slopes of
photoproduction samples is clearly larger than those of DIS samples. However
the model  assumes that the t-slope of the $\pi NN $ vertex is close to zero which
 is contradiction with the analyses of  the antiquark distributions in nucleons, see a review in
 \cite{Kumano:1997cy}. Also the model leads to much smaller rates of the neutron production 
 for the dijet trigger.
 
%-------------------------------------------------

\subsection{Inclusive diffractive measurements}

%-------------------------------------------------

\begin{figure}
\centerline{
  \includegraphics[height=.43\textheight]{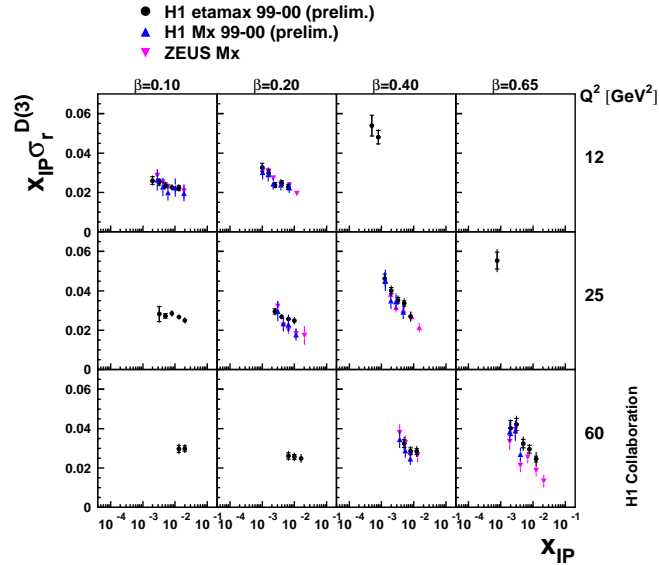}
}
  \caption{Comparisons of the diffractive reduced cross sections obtained using
	the LRG and $M_X$ methos with 1999-2000 data and with measurements
	from the ZEUS Collaboration.}
\label{fig_compH1LRGMX}
\end{figure}

H1 has presented the measurements of diffractive reduced cross sections covering
a wide kinematic range and obtained with the proton-tagging method\cite{H1FPS},
the LRG method\cite{H1LRG,H1Mx} and the $M_X$ method\cite{H1Mx}. 
Note that the reduced cross section is equal to the diffractive structure function
$F_2^{D(3)}$ except for the largest values of the inelasticity variable $y$ , where a correction
taking into account the effect of the longitudinal
component has to be considered. 
On Fig.~\ref{fig_compH1LRGMX}, the H1 experiment\cite{H1Mx} has  compared cross section values
based on  a LRG selection and a $M_X$ method, all together compared with ZEUS published results~\cite{zeusf2d} :
in the range of the comparison (medium $M_X$ vlaues), a good agreement is observed
for all these data sets and methods.

%-------------------------------------------------

\subsection{Semi-inclusive diffractive measurements}

%-------------------------------------------------

The QCD factorization theorem predicts that the same DPDFs should describe
both inclusive DIS diffraction and the semi-inclusive processes like dijet or charm production. 
The analyses of the data clearly show that the gluon
density largely dominates the quark content of the diffractive exchange~\cite{H1LRG,Royon,Watt}.

Several issues have been discussed on such procedures~\cite{Royon,Watt} concerning the
proper Regge factorisation implementation and relative importance of the leading twist and higher twist contributions in particular due to the interaction of the small size dipoles (which dominate in high $Q^2$ exclusive  diffraction).

In particular, the approach described in Ref.~\cite{Watt} does not
assume Regge factorization and 
suggests that the collinear factorization theorem,
though valid asymptotically in diffractive DIS, has important
modifications at the energies
relevant at HERA and predicts a significantly larger value of $\alpha_{\pom}(0)$ in diffraction than in soft processes.

The new data reported at the meeting seem to suggest
that the effective Pomeron trajectory for diffraction has an intercept $\alpha_{\pom}(0)= 1.118 \pm 0.008 {\rm (exp.)} ^{+0.029}_{-0.010} {\rm (model)}$
which is close to the one for the nonperturbative Pomeron.

The comparison 
of semi-inclusive final states,
as diffractive dijets at HERA~\cite{mozer,zeusdijet}
 and inclusive  data provides now a convincing confirmation of the QCD factorization for $\beta\le 0.4$. For larger $\beta$ the 
uncertainties in extraction of the gluon DPDF appear to be  too large to reach definite conclusions about consistency of the gluon densities extracted from DIS and from semi-inclusive processes.

\begin{figure}
  \hspace{2mm}
  \includegraphics[height=.06\textheight]{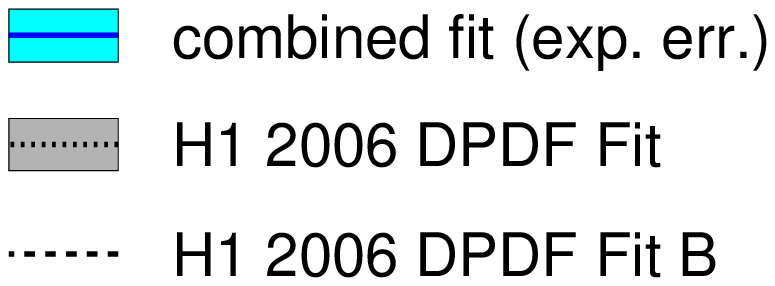}
  \hspace{4mm}
  \includegraphics[height=.35\textheight]{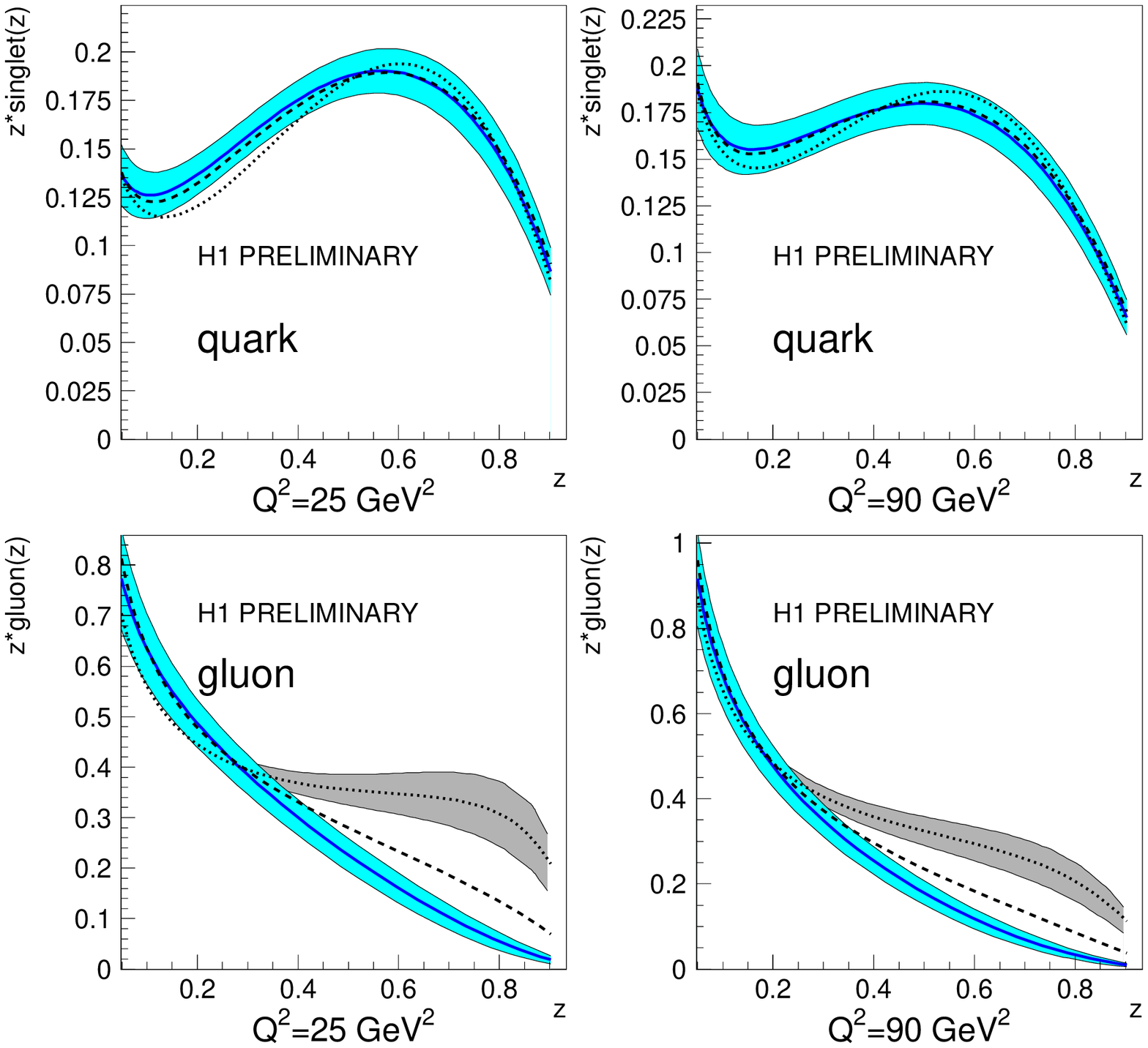}
  \caption{The diffractive singlet density (top) and diffractive gluon density
	(bottom) for two values of the hard scale $\mu$:25 GeV$^2$ (left) 
	and 90 GeV$^2$ (right). The blue line indicates the combined fit,
	surrounded by the experimental uncertainty band in light blue. 
 	For comparison,the two dashed lines show two fit results from the inclusive cross sections alone.
           }	
  \label{fig:H1Fit_PDF}
\end{figure}

\begin{figure}[tp]
\epsfxsize=1.0\textwidth
%\centerline{\epsfig{figure=BLESS_rjj1_dpemc_fit.eps,width=0.33\hsize}
%            \epsfig{figure=BLESS_norm_fbc_rjj_w_pomwig.eps,width=0.33\hsize}
%            \epsfig{figure=BLESS_cdf+h1fit2_rjj.eps,width=0.33\hsize}}
\centerline{\epsfig{figure=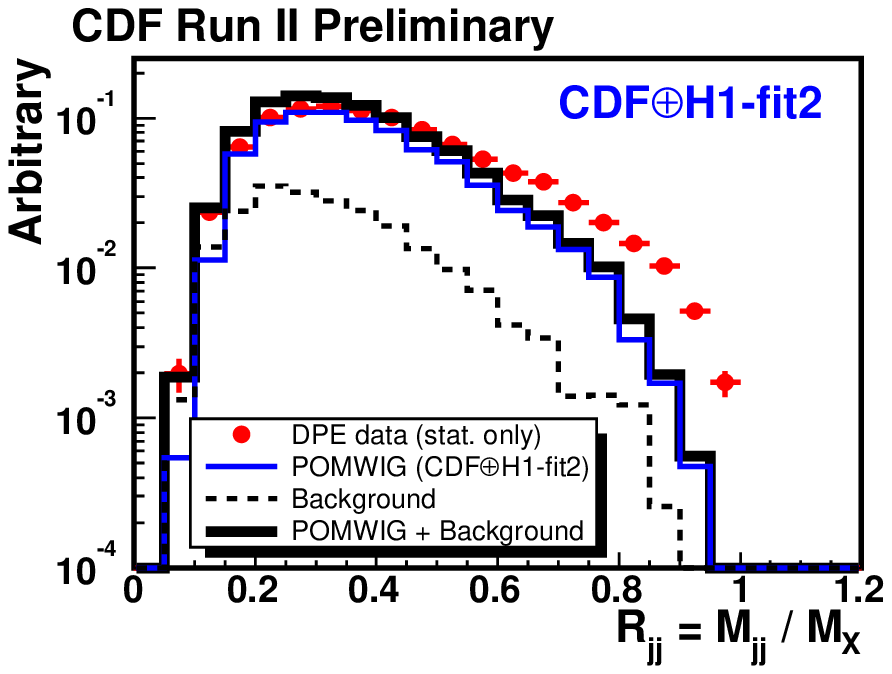,width=0.33\hsize}
            \epsfig{figure=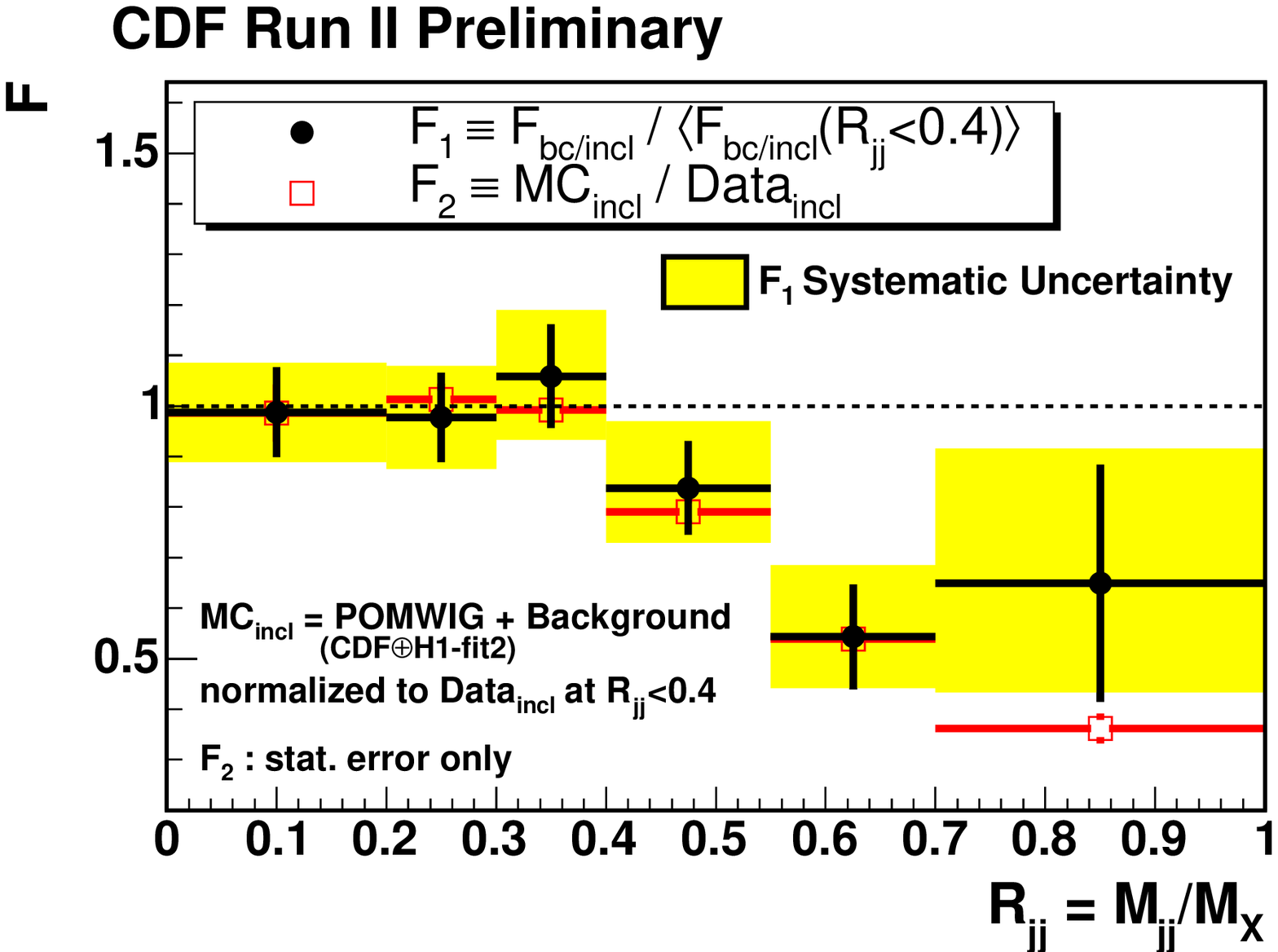,width=0.33\hsize}
            \epsfig{figure=bless_cdf+h1fit2_rjj.eps,width=0.33\hsize}}
\caption{\label{fig:excl}
{\em Left}: dijet mass fraction in DPE data (points) and best fit (solid) obtained from POMWIG MC events (dashed) and exclusive dijet MC events (shaded);
{\em Center}: normalized ratio of heavy flavor jets to all jets as a function of dijet mass fraction.
{\em Right}: $R_{jj}$ distribution for the data (points) and POMWIG MC prediction (thick histogram),
composed of DPE dijet events (thin) and non-DPE events (dashed). Data and MC are normalized to the same area.
}
\end{figure}

Fig~\ref{fig:H1Fit_PDF} presents the gluon distribution of the diffractive exchange extracted
by combining the inclusive and dijet measurements from the H1 experiment, which leads to 
a much better constraint on the gluon density at large momentum fraction.
At TEVATRON also, dijet events in 
double Pomeron exchange (DPE) show a strong sensitivity to diffractive gluon density
~\cite{Royon,ourpaper} and new experimental results have been presented~\cite{cdf} in terms of the dijet mass fraction.
This variable  ($R_{jj}$) is defined as the dijet invariant mass ($M_{jj}$) divided by the mass of the entire system $M_X$, calculated using all available energy in the calorimeter.
If jets are produced exclusively, $R_{jj}$ should be equal to one. 
Data are compared with MC expectations on Fig.~\ref{fig:excl}.
At large $R_{jj}$ values, the excess of events in the data with respect to inclusive DPE dijet production, which is described by POMWIG\cite{pomwig} MC,
is well accounted for by the DPEMC\cite{dpemc}
MC sample of exclusive events (Fig.~\ref{fig:excl}, left).

Another process illustrated in Ref.~\cite{cdf} concerns
exclusive diphoton events, $p\overline{p}\rightarrow p\gamma\gamma\overline{p}$.
The final state is cleaner than in exclusive dijet production as hadronization effects are absent,
but the expected cross section is smaller.

% <------------------------------------------------------------------------

%-------------------------------------------------

\section{Exclusive final states}

%-------------------------------------------------

The dynamics of diffractive interactions can also be studied through 
exclusive vector meson ($\rhoz, \omega, \jpsi, ... $) 
and photon production, $ l + N \longrightarrow l + V + Y$, 
where $Y$ is defined as in previous sections as an elastically scattered nucleon or a low-mass dissociative
state. At low transverse momentum transfer
at the nucleon vertex, the photoproduction of \rhoz, $\omega$ and $\phi$
mesons is characterized by a ``soft'' dependence of their cross-sections
in the $\gamma p$ center-of-mass energy, $W$. This can be interpreted 
in the framework of Regge theory as due
to the exchange of a ``soft'' Pomeron (\pom) resulting in an energy dependence
of the form $\rm{d}\sigma / \rm{d} t \ \propto \ W^{4(\apom(t)-1)}$, 
where the Pomeron is parametrised by a Regge trajectory.

The H1 collaboration has performed a new cross section measurement of the exclusive \rhoz
photoproduction, using a sample of more than 240k events recorded in the year 2005.
This measurement has been done in the kinematic domain $20<W<90$~GeV and $|t|<3$~GeV$^2$,
leading to a determination of the $W$ dependence of the cross section
in eight bins of $t$~\cite{olson}.

%%%

Assuming that the Pomeron trajectory is linear they find:
$\apom(t) = \apom(0) + \aprim t \simeq 1.093 + 0.116 \ t$,
which is corresponds to a factor $\sim 2$ smaller \aprim than measured in  the $pp$ elastic scattering. However the data do not exclude a possibility that \aprim is about the same for small t, but that effective Pomeron trajectory is nonlinear. Note also that the data use a Regge factorization model to subtract the inelastic background. This assumption is expected to be violated at sufficiently large t where pQCD contribution to inelastic diffraction becomes important.

The QCD factorization theorem is valid for production of vector mesons by a large $Q^2$ longitudinally polarized photons, and for production of  mesons build off heavy quarks \cite{Collins}. Also, it is often assumed  that perturbative QCD is valid if the "hard scale"
 is provided by the momentum transfer \ttra\, or highly virtual transversely polarized photon
 or of the vector meson mass, perturbative QCD is expected to
apply.  In this approximation,
diffractive vector meson production can be seen in the nucleon
rest frame as a sequence of three subprocesses well separated in time:
the fluctuation of the exchanged photon in a \qqbar\ pair, the hard 
interaction of the \qqbar\ pair with the nucleon via the exchange of
(at least) two gluons in a color singlet state, and the \qqbar\ pair
recombination into a real vector meson. This approach results in a stronger
rise of the cross section with $W$ than for soft processes, which reflects 
the rise at small $x$ of the gluon density in the nucleon. 
 Furthermore, to take into account the  skewing effect, i.e. 
the difference between the proton momentum fractions carried by the two 
exchanged gluons, one has to consider generalized parton distributions 
(GPDs). All these aspects have been covered with new results 
and are summarised in the following.

The H1 collaboration has  finalized a determination of the 
diffractive production of \rhoz mesons at large 
$|t|$ ($1.5<|t|<10.0$~GeV$^{2}$), using data taken
during the year 2000, with an integrated luminosity of 20.1~pb$^{-1}$~\cite{olson}.
Several spin density matrix elements (SDME), which carry information
on the helicity structure of the production amplitudes, have been
extracted from the production and decay \rhoz\ angular distributions. 
The data indicate a violation of the {\it s}-channel helicity conservation (SCHC),
with contributions from both single and double helicity flip being observed.
This observations contrast to what was observed for high \ttra\ \jpsi\ production and 
 is generally attributed to differences
in the wave function between $\rho$ and \jpsi. 
The cross section, differential in $t$, for  $1.5<|t|<10.0$~GeV$^{2}$, 
is also in  reasonable agreement with
pQCD models, inspired by the  BFKL phenomenology.
 However these models are based on the leading order BFKL approximation
 which is known to differ strongly from resummed BFKL. Also none 
 of the available models  is able to explain behavior of all spin observables.  

\begin{figure}
  \centering
  \includegraphics[width=0.33\textwidth]{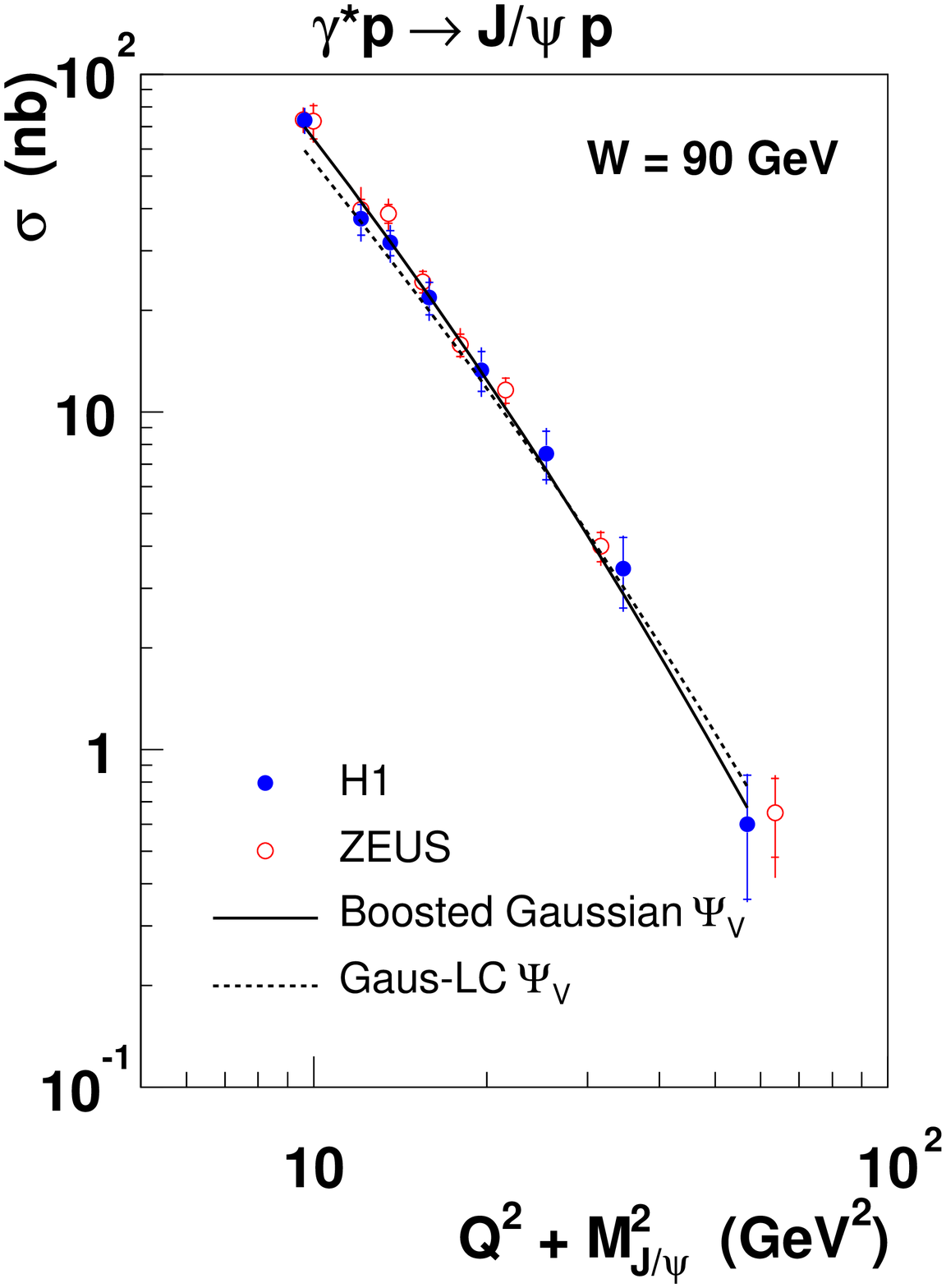}%
  \includegraphics[width=0.33\textwidth]{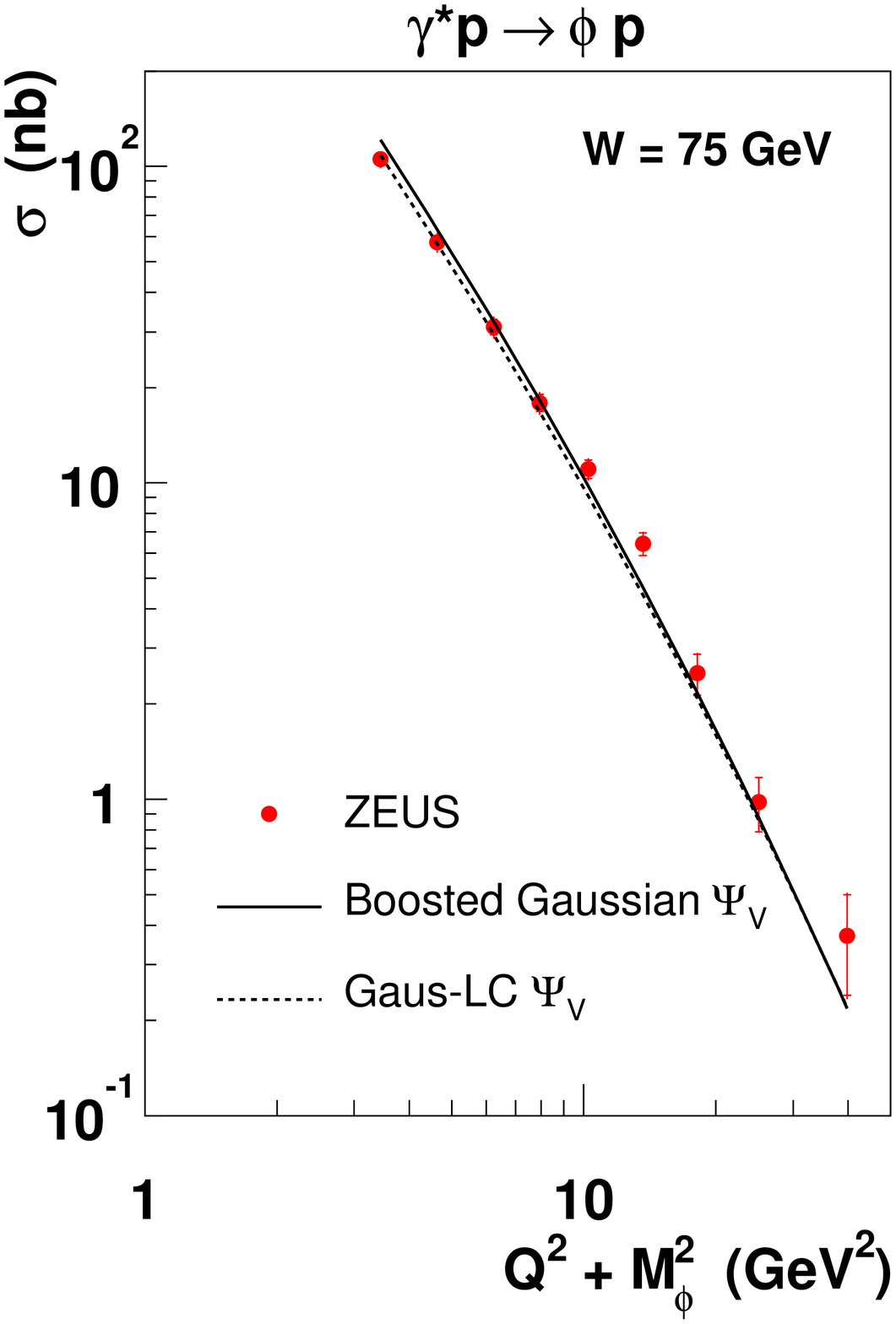}%
  \includegraphics[width=0.33\textwidth]{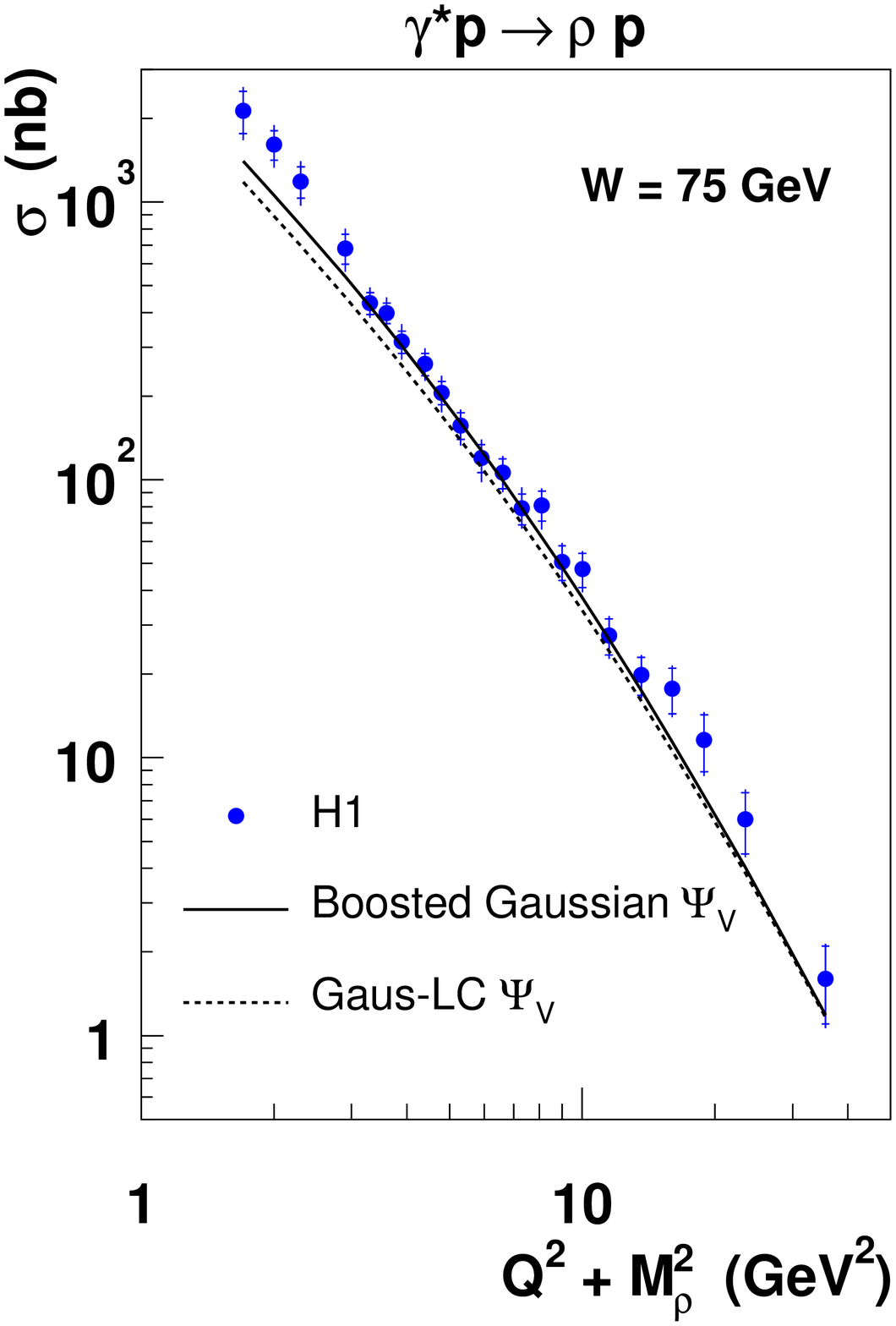}
  \caption{Total vector meson cross section $\sigma$ vs.~$(Q^2+M_V^2)$ compared to predictions from the b-Sat model using two different vector meson wave functions.}
  \label{fig:crossq}
\end{figure}

\begin{figure}
  \centering
  \includegraphics[width=0.33\textwidth]{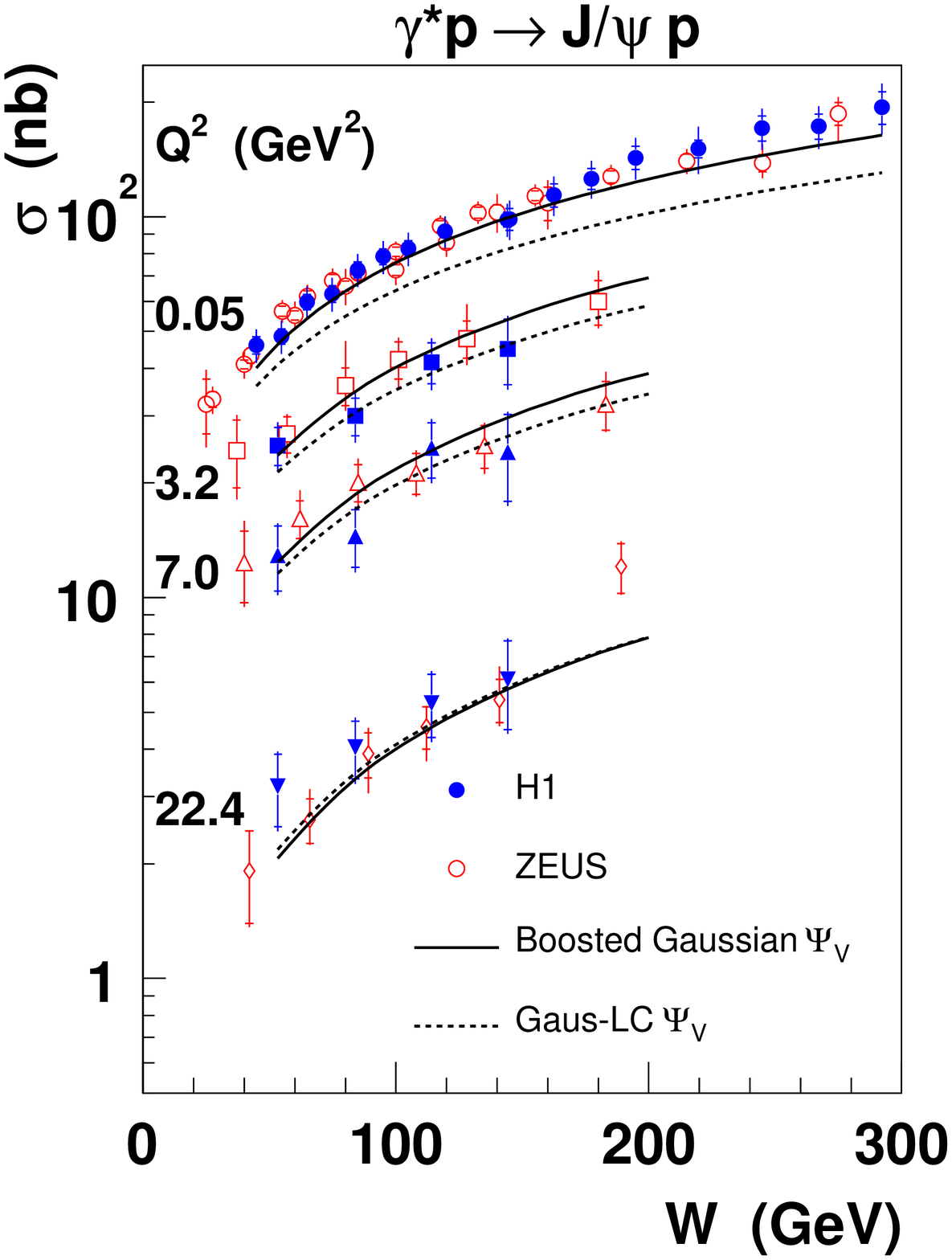}%
  \includegraphics[width=0.33\textwidth]{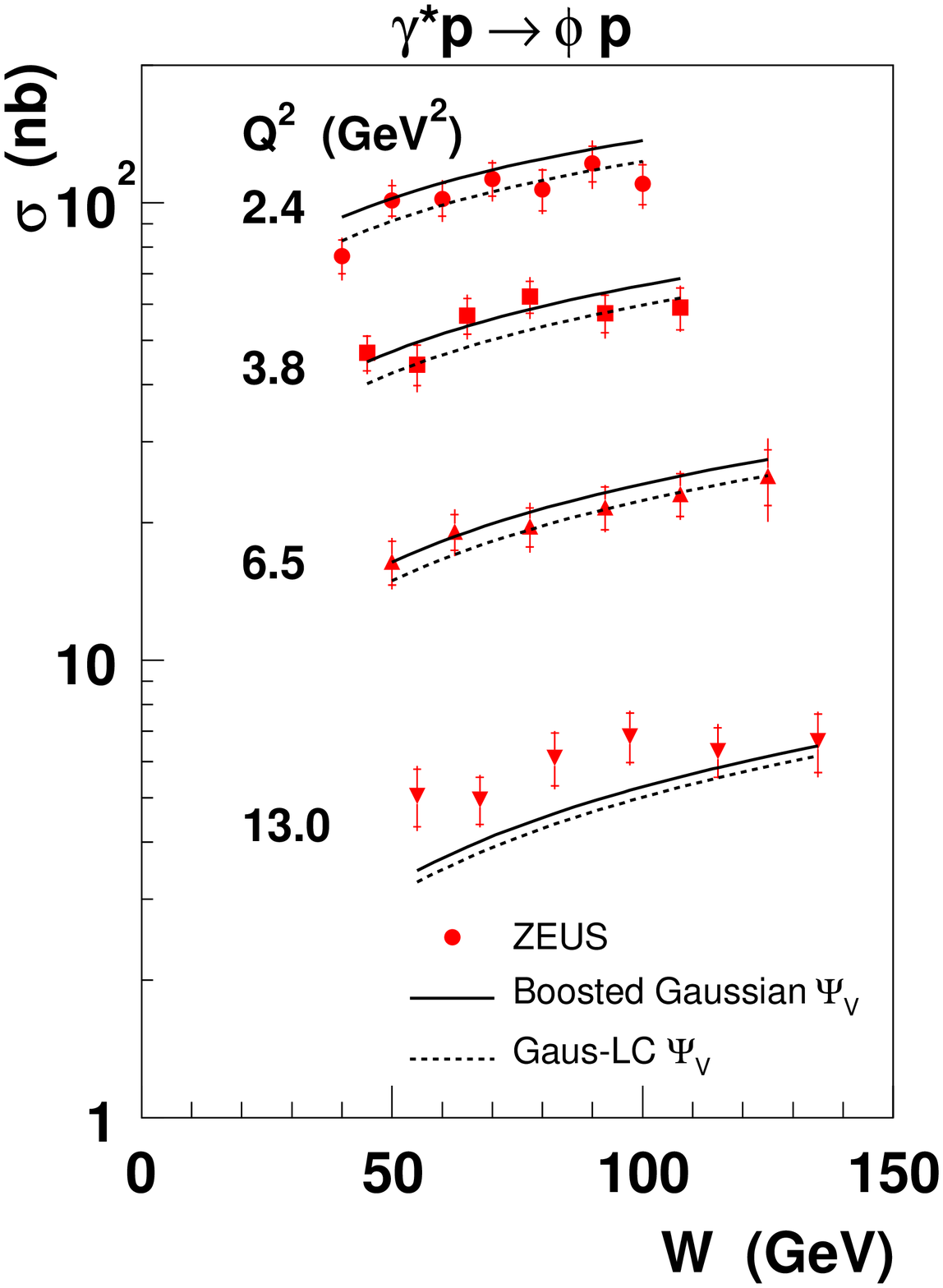}%
  \includegraphics[width=0.33\textwidth]{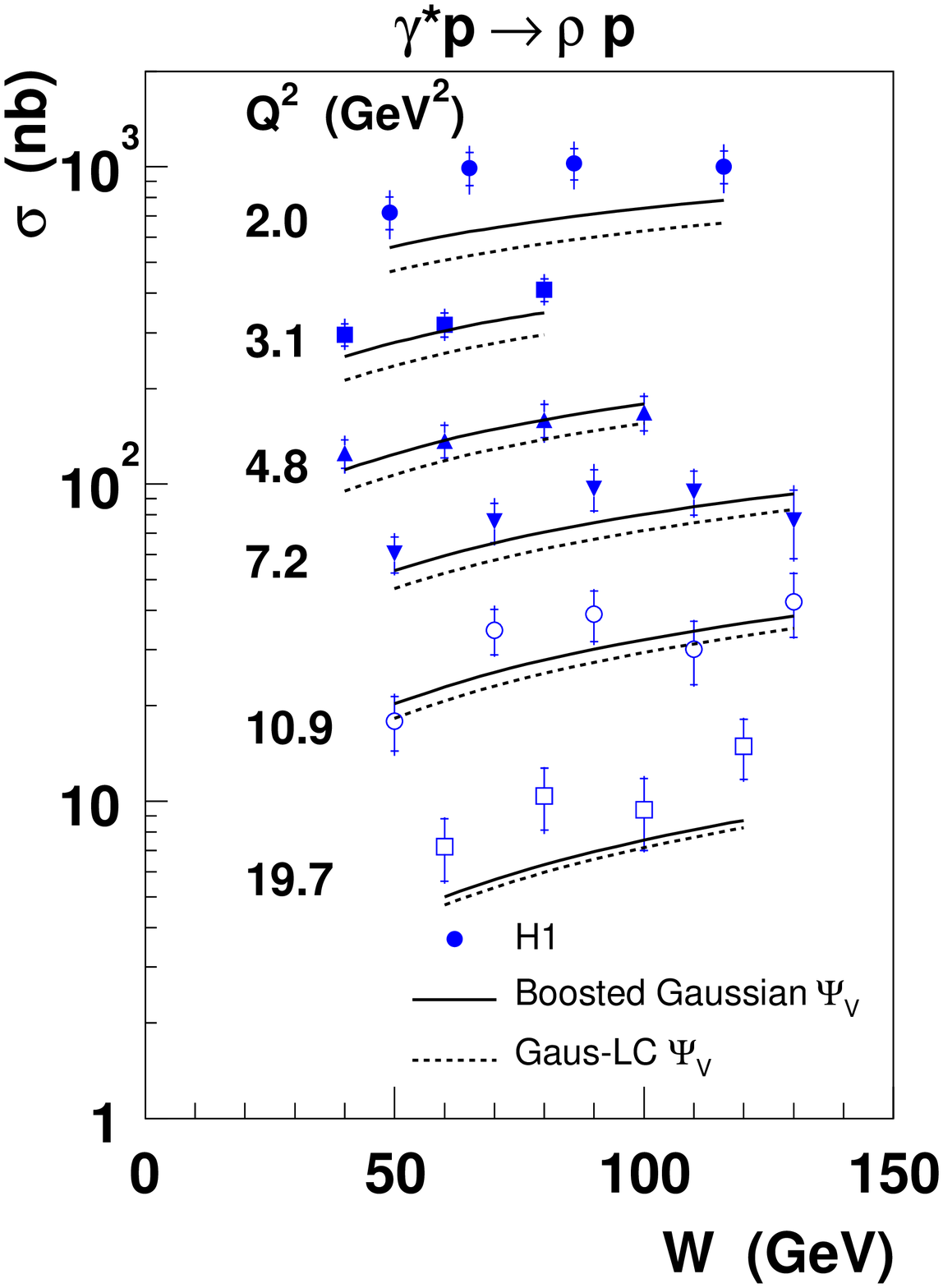}
  \caption{Total vector meson cross section $\sigma$ vs.~$W$ compared to predictions from the b-Sat model using two different vector meson wave functions.}
  \label{fig:crossw}
\end{figure}

New developments on dipole models based on the saturation approach to describe
the dipole-proton cross section have been presented~\cite{Kowalski}. They provide a very
good fit for the cross sections of all exclusive processes (in the DIS regime),
differential in $Q^2$, $W$ or $t$ (see Fig.~\ref{fig:crossq} 
and Fig.~\ref{fig:crossw}).

\begin{figure}[tb]
\begin{center}
\includegraphics[width=7cm]{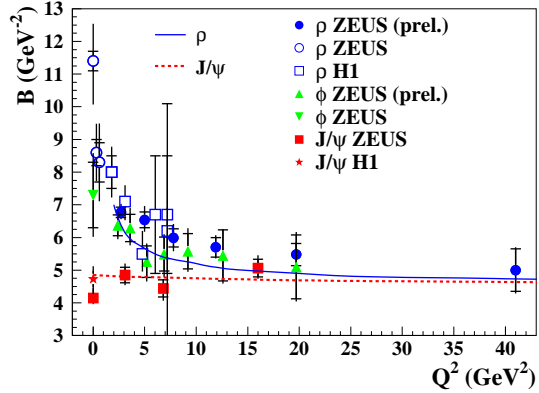}
\caption{\label{b-slopes} The logarithmic slope of the $t$ dependence
  at $t=0$ for different meson production channels.}
\end{center}
\end{figure}

One of the key perturbative QCD predictions was universality of the slopes 
(defined  on  the exponential fit to the differential cross section:  $d\sigma/dt \propto
\exp(-b|t|)$ at small $t$) in the leading twist limit \cite{Brodsky:1994kf}. 
The data shown in Figure~\ref{b-slopes} are consistent with onset of such universality.
The dipole model is well suited to estimate the onset of the scaling regime by estimating contribution to the slope of the finite size of the $q\bar q$ dipole.
The rate of   convergence of the slopes is consistent with predictions of \cite{FKS97}, see the lines in Fig.~\ref{b-slopes}.

\vspace*{-0.3cm}

A Fourier transform from momentum
to impact parameter space readily shows that $b$ is related to the
typical transverse distance between the colliding objects, see \cite{diehl} and references therein.  
At high scale, the $q\bar{q}$ dipole is almost
point-like, and the $t$ dependence of the cross section is controlled by
the $t$ dependence of the generalized gluon distribution, or in
physical terms, by the transverse extension 
of the gluons in the  proton for a given $x$ range.

\begin{figure}[ht]
  \begin{minipage}{0.48\textwidth}
    \includegraphics[clip,width=0.98\textwidth]{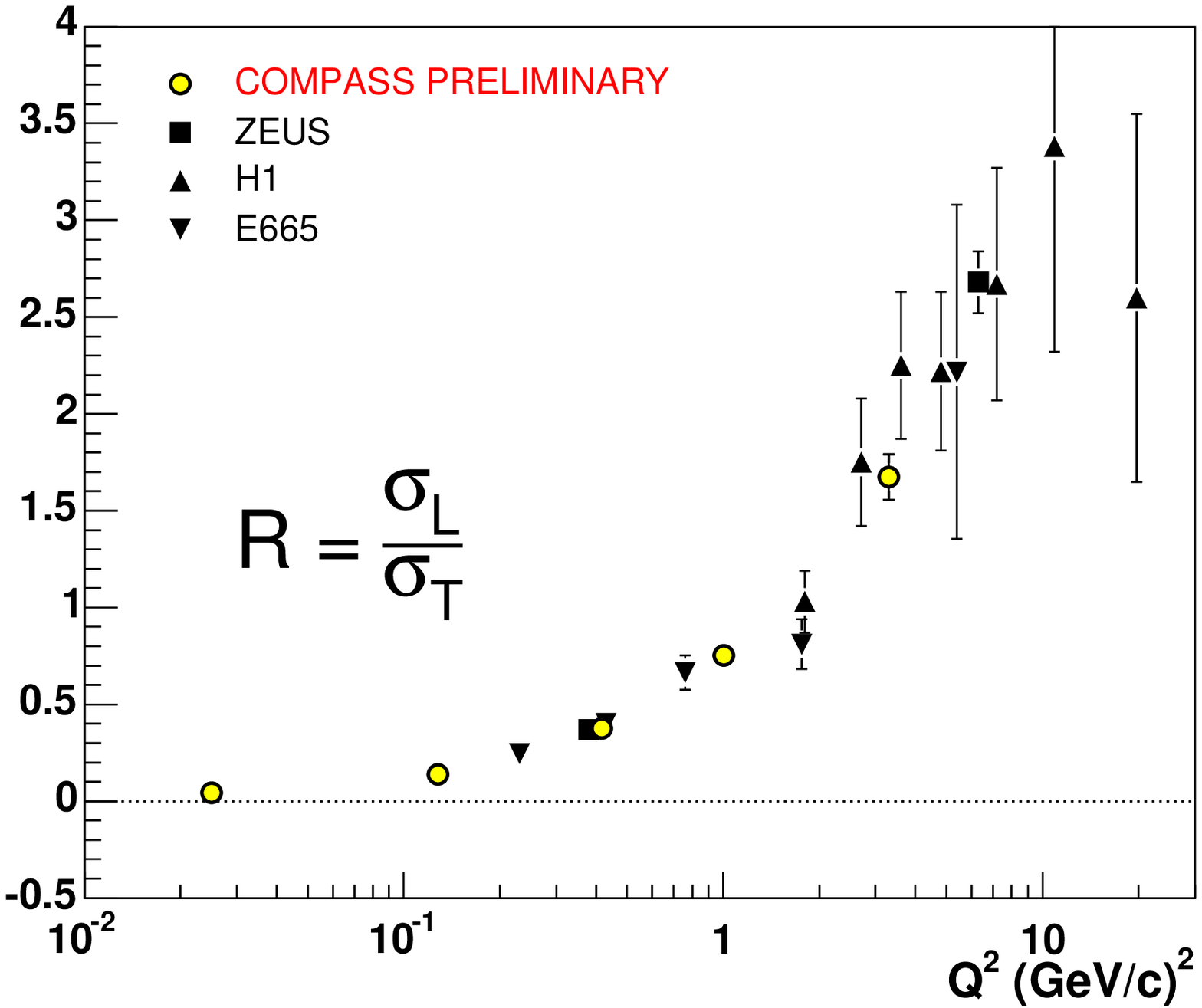}
    \caption{$Q^2$ dependence of the ratio R between the longitudinal
      and transverse cross section, derived from elastic \rhoz\ mesons
      production as measured by COMPASS.}  
    \label{fig0}
  \end{minipage}\hfill
  \begin{minipage}{0.48\textwidth}
    \includegraphics[clip,width=0.98\textwidth]{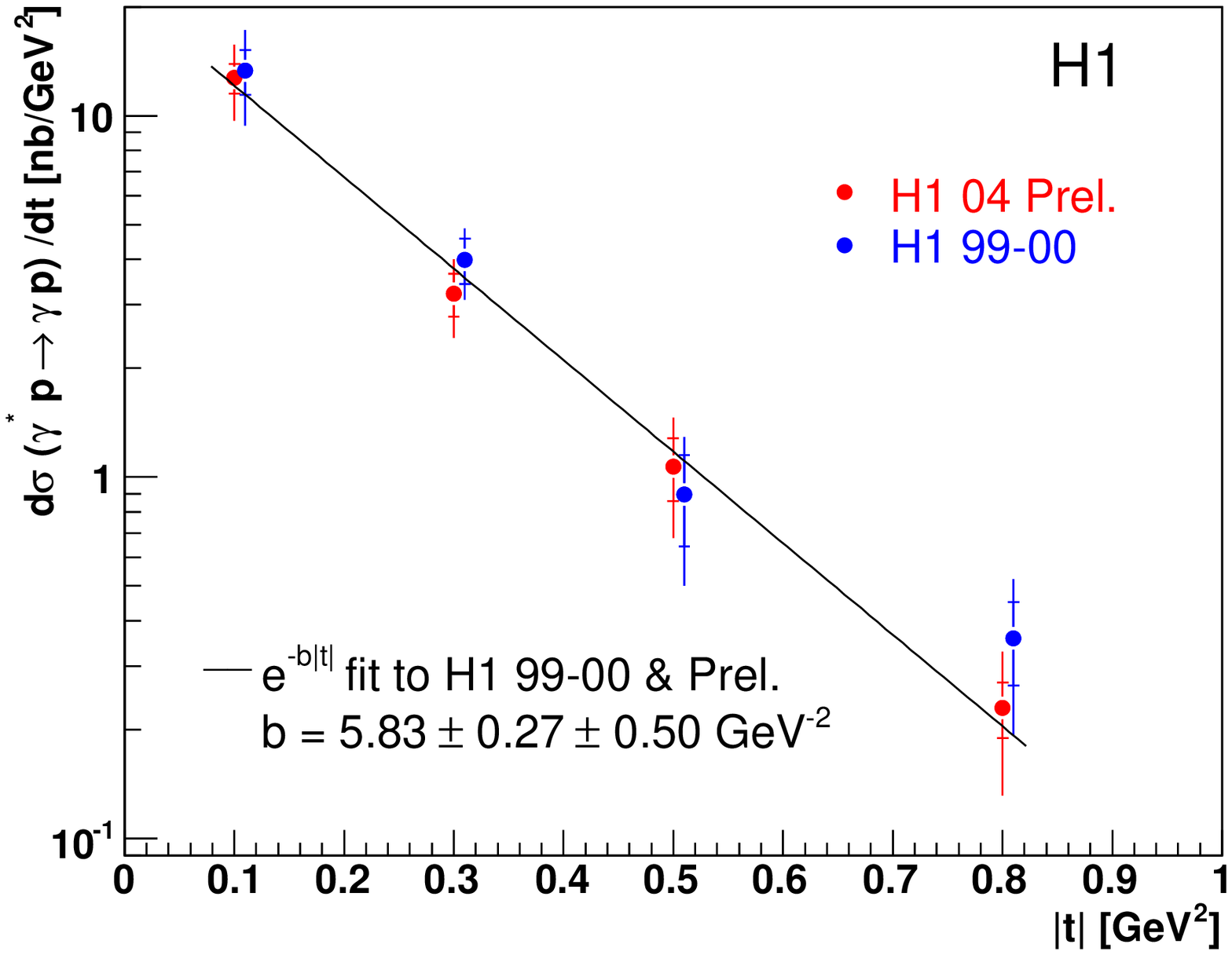}
    \caption{DVCS cross section differential in $t$ at $Q^2=8$~GeV$^2$.}
    \label{fig2}
  \end{minipage}
\end{figure}

The COMPASS experiment has presented a study of the 
diffractive elastic leptoproduction of \rhoz\ mesons, 
$\mu + N \longrightarrow \mu + \rhoz + N$, where $N$ is a quasi-free
nucleon from any of the nuclei of their polarized target, at 
$<W> = 10$ \gev\ for a wide range of \qsq, $0.01 < \qsq < 10$ \gevsq~\cite{nicole}.
The COMPASS data provide a large statistics which allows to extend 
the previous measurements  of the
\rzqzz\ matrix element towards low \qsq.  
Then, if one assumes SCHC between
the exchanged photon and the \rhoz\ meson, one can obtain the  ratio $R$ 
between the longitudinal ($\sigma_L$) and the transverse ($\sigma_T$) 
cross sections (see Fig.~\ref{fig0}). A complementary analysis has been presented
by HERMES ~\cite{hermesnew} with the measurements of the $Q^2$ and $t$
dependences for the 15 unpolarized SDMEs (from $0.7$~GeV$^2$ till $5$~GeV$^2$
and $|t|<0.4$~GeV$^2$).

Information about GPDs in lepton nucleon scattering can be provided 
by measurements of exclusive processes in which the nucleon remain intact.
A complete overview of the topic has been presented in Ref.~\cite{diehl}.
In particular, the simplest process sensitive to GPDs is Deeply Virtual Compton
Scattering (DVCS), i.e. exclusive photon production off the proton 
$\gamma^* p \longrightarrow \gamma p$ at small \ttra\ but
large \qsq, which is calculable in perturbative QCD.
Such a final state also receives contributions from the purely
electromagnetic Bethe-Heitler process, where the photon
is radiated from the lepton. The resulting interference term
in the cross section vanishes as long as one integrates over
the azimuthal angle between the lepton and the hadron plane. 
It is then possible to extract the DVCS cross section by
subtracting the Bethe-Heitler contribution, as done by H1.

A new high statistics analysis of DVCS has been performed by the
H1 experiment in the kinematic region $4<\qsq<80$
\gevsq, $30<W<140$ \gev\ and $\ttra<1$ \gevsq, using data taken during the 
year 2004. The $\gamma^* p 
\longrightarrow \gamma p$ cross section has been measured as a
function of \qsq\ and as a function of $W$~\cite{roland}. The $W$ dependence
can be parametrised as $\sigma \propto W^{\delta}$, yielding
$\delta \simeq 1.0$ at $\qsq=8$ \gevsq, i.e.
a value similar to \jpsi\ production indicating the presence
of a hard scattering process.
 For the first time at HERA II, the DVCS
cross section has been measured differentially in $t$ 
  and the observed fast decrease with \ttra
can be described by the form $e^{-b\ttra}$ with
$b=5.83 \pm 0.27 \pm 0.50$ \gevsq\ at $\qsq=8$ \gevsq (see Fig.~\ref{fig2}).The observed t-slope of DVCS is substantially larger than the gluon GPD slope. 

This maybe due to a larger radius of the quark transverse distribution at intermediate $Q^2$ due to the pion cloud contribution\cite{Weiss}. This effect would be amplified by a factor of $\sim 2$ for the DVCS slope as the gluon GPD contribution as calculated in the NLO \cite{Martin-Freund} enters with negative sign with weight $\sim 0.5$.

% GPDs Compass

As discussed above, experiments at HERA collider are well 
designed to study mainly the gluon GPDs at very small \( x_{Bj} \)
(\( \leq 10^{-2} \)). The experimental program using
COMPASS at CERN (at 100 and/or 190 GeV) will enlarge the kinematical
domain to a large range of \( Q^{2} \) and \( x_{Bj} \) (1.5 \( \leq Q^{2}\leq  \)
7 GeV\( ^{2} \) and 0.03 \( \leq x_{Bj}\leq  \) 0.25),
from the process $\mu p \rightarrow \mu' p'\gamma$. 
COMPASS is the unique place to measure the azimuthal distribution of
the Beam Charge Asymmetry which seems very promising to test the geometrical
interpretation of GPDs~\cite{heinsius}.
%The first one is based on
%a simple parametrization of the GPDs: \( H^{f}(x,\xi ,t)=H^{f}(x,\xi ,0)F^{f}_{1}(t)/2 \)
%where \( F^{f}_{1}(t) \) represents the elastic Dirac form factor
%for the quark flavor \( f \) in the nucleon. The second one
%relies on the fact that the GPDs measure the contribution of quarks
%with longitudinal momentum fraction \( x \) to the corresponding
%form factor as is suggested by the sum rule: $
%\int ^{+1}_{-1}H^{f}(x,\xi ,t)dx=F^{f}_{1}(t)$.
%\begin{figure}
%{\centering \resizebox*{0.8\textwidth}{!}{\includegraphics{bca_q2_x_th1_fin.eps}} \par}
%\caption{\label{azymuth}Azimuthal distribution of the beam charge asymmetry
%measured at COMPASS at \protect\( E_{\mu }\protect \)= 100 GeV and
%\protect\( |t|\leq 0.6\protect \) GeV\protect\( ^{2}\protect \)
%for 2 domains of \protect\( x_{Bj}\protect \) (\protect\( x_{Bj}=0.05\pm 0.02\protect \)
%and \protect\( x_{Bj}=0.10\pm 0.03\protect \)) and 3 domains of \protect\( Q^{2}\protect \)
%(\protect\( Q^{2}=2\pm 0.5\protect \) GeV\protect\( ^{2}\protect \),
%\protect\( Q^{2}=4\pm 0.5\protect \) GeV\protect\( ^{2}\protect \)
%and \protect\( Q^{2}=6\pm 0.5\protect \) GeV\protect\( ^{2}\protect \))
%obtained in 6 months of data taking with a global efficiency of 25\%
%and with \protect\( 2\cdot 10^{8}\protect \) \protect\( \mu \protect \)
%per SPS spill (\protect\( P_{\mu ^{+}}=-0.8\protect \) and \protect\( P_{\mu ^{-}}=+0.8\protect \)) }
%\end{figure}

%-------------------------------------------------

\section{Towards LHC}

%-------------------------------------------------

In recent years, the production of the Higgs boson in diffractive 
$pp$ collisions has drawn more and more attention as a clean channel to 
study the properties of a light Higgs boson or even discover it. This is 
an interesting example of a new theoretical challenge: to adapt and apply the 
techniques for the QCD description of diffraction in $ep$ collisions to 
the more complex case of $pp$ scattering at the LHC.  
We have already shown in previous sections how first results 
on hard diffraction can be transmitted from HERA to TEVATRON.
The experimental interests 
to study diffractive processes at the LHC, in connection
with the  proposal to add forward proton detectors,
have grown in parallel with the theory.
Various aspects of physics with forward proton tagging
at the LHC have been under discussion in our working group~\cite{Royon,cox}.
In Ref.~\cite{cox,martin}, the
unique physics potential of forward proton tagging at 420m at the LHC
have been presented and discussed.

In Ref.~\cite{Royon} inclusive and 
exclusive models have been discussed. 
Experimentally at LHC, for exclusive events, the full energy available
in the center of mass is used to produce the heavy object (dijets, Higgs,
diphoton, $W$...).  For inclusive events,
only part of the available energy is used to produce the heavy
object diffractively. In Ref.~\cite{Royon}, it is assumed that the Pomeron is made of quarks
and gluons (with the gluon and quark densities deduced from the HERA measurements 
in shape and the normalisation from TEVATRON data). Then, a quark or a gluon from
the Pomeron is used to produce the heavy state. In this context, exclusive model appear
to be the limit where the gluon in the Pomeron is a $\delta$ distribution, with 
no Pomeron remnants for exclusive events.
It is shown~\cite{Royon} that this distinction is quite relevant for
experimental applications.

Theoretical calculations of exclusive processes were reviewed in \cite{martin}.
The survival probability due to the soft interactions of 0.026 was reported and it was argued that the perturbative effects suggested by Bartels et al \cite{Bartels}  are stongly overestimated. Predictions for a wide range of exclusive channels were presented which could be checked already at the TEVATRON collider. 
In \cite{Hyde-Wright} that the t-slope of the gluon GPD in the kinematics relevant for the calculation of the exclusive  Higgs production at LHC is substantially smaller than the one assumed by \cite{martin}. Using such a smaller slope would result within the model of \cite{martin} to reduction of the gap survival probability by a factor of $\sim 3$.  Also it was suggested in \cite{Hyde-Wright} that by
 measuring the ``diffraction pattern,'' of the $p_t$ dependence of the scattered  protons one can 
perform detailed tests of the interplay of hard and soft interactions, and 
even extract information about the gluon GPD in the proton from the data.

Perspectives of the studies of hard photon- proton (nucleus) interactions at LHC using ultraperipheral collisions of protons (nuclei) with nuclei were reviewed in \cite{Strikman}. It was demonstrated that such measurements would allow to extend the measurements of exclusive production of heavy mesons as compared to HERA to $W\sim 1 TeV$ and allow to measure the gluon densities 
down to $x\sim 10^{-4}$ for $p_t \sim 6 $ GeV. The same measurements would allow to check the prediction that even for large virtualities probability of the diffractive events for which nucleus would remain intact will be of the order 0.2.

%-------------------------------------------------

\section{Conclusions}

The findings presented at the session  provided further constrains on the interplay between soft and hard physics in the diffraction processes. The forthcoming results from the data analyses of the recent data and the future run at lower energy will allow to move further in resolving open questions such as the dependence of 
$\alpha^{\prime} $ on the hardness scale, contribution of higher twist processes in the inclusive and exclusive diffraction, energy dependence of the exclusive processes at high resolution scales.

Coordinators would like to thank the participants of the session  as well as J.Bartels, A.Freund and M.Ryskin for numerous contributions and inputs.

%-------------------------------------------------

%-------------------------------------------------

\end{document}